\documentclass[aps,prd,superscriptaddress,floatfix,nofootinbib,showpacs,twocolumn]{revtex4}       

\usepackage{amsmath,amssymb}       

\usepackage{color}       
\usepackage[dvips]{graphicx}       

\newcommand{\lsim}{\mathrel{\mathop{\kern 0pt \rlap       
  {\raise.2ex\hbox{$<$}}}    
  \lower.9ex\hbox{\kern-.190em $\sim$}}}    
\newcommand{\gsim}{\mathrel{\mathop{\kern 0pt \rlap    
  {\raise.2ex\hbox{$>$}}}    
  \lower.9ex\hbox{\kern-.190em $\sim$}}}    
\newcommand{\sigmav}{\langle \sigma_{\rm ann} v \rangle}

\newcommand{\beq}{\begin{equation}}    
\newcommand{\eeq}{\end{equation}}    
\newcommand{\bea}{\begin{eqnarray}}    
\newcommand{\ena}{\end{eqnarray}}

\begin{document}    
    
    
\title{Upper bounds on signals due to WIMP self--annihilation: comments on the case of the    
synchrotron radiation from the galactic center and the WMAP haze}

\author{A. Bottino}    
\affiliation{Dipartimento di Fisica Teorica, Universit\`a di Torino \\    
Istituto Nazionale di Fisica Nucleare, Sezione di Torino \\    
via P. Giuria 1, I--10125 Torino, Italy}    
    
\author{F. Donato}    
\affiliation{Dipartimento di Fisica Teorica, Universit\`a di Torino \\    
Istituto Nazionale di Fisica Nucleare, Sezione di Torino \\    
via P. Giuria 1, I--10125 Torino, Italy}    

\author{N. Fornengo}    
\affiliation{Dipartimento di Fisica Teorica, Universit\`a di Torino \\    
Istituto Nazionale di Fisica Nucleare, Sezione di Torino \\    
via P. Giuria 1, I--10125 Torino, Italy}    

\author{S. Scopel}    
\affiliation{Korea Institute for Advanced Study \\    
 Seoul 130-722, Korea}

\date{\today}    
    
\begin{abstract}    
Two recent papers reconsider the possibility that the excess of    
microwave emission from a region within $\sim 20^0$ of the galactic center    
(the {\it  WMAP haze}),  measured by WMAP,  can be due to    
the synchrotron emission originated by neutralino self-annihilation; on the basis    
of this possible occurrence, also upper bounds on the    
neutralino self-annihilation cross--section are suggested. In the present note, we show that in the common case of thermal WIMPs in a standard cosmological model, when    
the rescaling of the galactic WIMP density is duly taken into account for subdominant    
WIMPs, the upper bound applicable generically to {\it any} signal due to self-conjugate     
WIMPs is more stringent than the ones obtained from analysis of the WMAP haze. 
We also argue that an experimental upper bound,  which    
can compete with our generic upper limit, can rather be derived from measurements of cosmic antiproton fluxes,  for some values of the parameters of the astrophysical propagation model. 
Finally, we comment on the possible impact of our generic upper bound  
on the interpretation of the WMAP haze in terms of thermal neutralinos in a standard 
cosmological scheme. 

\end{abstract}    
    
    
\maketitle


Synchrotron radiation, emitted by positrons and electrons created    
by self--annihilation of WIMPs in the galactic center, has long since been    
analyzed  as a potentially interesting signature for    
particle dark matter in our Galaxy.     
In Refs. \cite{fink,hfd} it has been suggested  that the synchrotron    
emission due to WIMPs can also be responsible for an excess of    
microwave emission from a region within $\sim 20^0$ of the galactic center    
(the {\it  WMAP haze}), measured by WMAP \cite{wmap06}.    
In Ref. \cite{hooper} this possible connection of the WMAP haze    
with WIMP annihilation in the galactic center has been    
reconsidered, with the purpose of putting upper bounds on the    
neutralino self-annihilation cross--section. The analysis of Ref.    
\cite{hooper} is carried out in terms of various annihilation    
channels and dark matter density profiles.     

In the present note we address some of the previous aspects in the usual
framework of thermal WIMPs in a standard cosmological model.
First we recall that a generic upper bound for indirect signals due to
self--interaction
of {\it any} thermal self--conjugate WIMP can be derived in a
model--independent way. Then we show that an experimental upper bound  which
can compete with the generic upper limit and is conservatively stronger
than the one obtained from the WMAP haze
 is derivable from measurements of cosmic antiproton fluxes, for some
 values of the parameters of the astrophysical propagation model.
Finally, we discuss
the relevance of the generic upper limit
for an interpretation of the WMAP haze in terms of  neutralino
self--annihilation in the galactic center. In what follows, we use the
subscript $\chi$ to denote properties related to our generic
self--conjugate WIMP (not necessarily a neutralino).
    
The first point was discussed in Refs. \cite{bffms,lathuile} and    
further developed in Ref. \cite{bdfs}. The reasoning goes as    
follows. Any signal due to WIMP self-annihilation is proportional    
to $\rho_{\chi}^2(\vec{r}) \sigmav_0$, where $\rho (\vec{r})_{\chi}$ is the    
WIMP galactic distribution and $\sigmav_0$ is the average, over    
the Galactic velocity distribution, of the WIMP annihilation    
cross-section multiplied by the relative velocity. For    
convenience, we rewrite $\rho(\vec{r})_{\chi}$ as    
$\rho(\vec{r})_{\chi} = \rho(\vec{r}) \xi$, where $\rho(\vec{r})$ denotes the    
total dark matter galactic distribution and $\xi$ the fractional    
part due to the WIMP $\chi$. Thus, the signal depends on the    
specific physical properties of $\chi$ through the quantity    
$\xi^2  \langle\sigma_{\rm ann} v\rangle_0$, and not simply through    
$\langle\sigma_{\rm ann} v\rangle_0$.    
    
    This is a crucial point, which can be elucidated by    
employing the rescaling recipe of Ref.  \cite{gst} for obtaining the    
factor $\xi$, {\it i.e.} by setting    
 $\xi = \min[1,\Omega_{\chi} h^2/(\Omega_{\rm CDM}h^2)_{\rm min}]$,    
 where $(\Omega_{\rm CDM}h^2)_{\rm min}$ is the minimal amount    
 of cold dark matter in the Universe.    
We recall that this factor $\xi$ has the effect of rescaling the    
WIMP galactic density, when its average relic density in the    
Universe is below the minimal amount required for cold dark matter    
on the basis of cosmological observations (in other words, when the    
generic candidate $\chi$ represents only a subdominant component of    
cold dark matter).    
    
 Once the factor $\xi$ is introduced by the rescaling recipe    
 of Ref.  \cite{gst}, one can readily    
 understand some relevant properties of    
 $\xi^2  \langle\sigma_{\rm ann} v\rangle_0$ as a function of the set 
$\eta$ of parameters of the particle physics model which describes our    
generic cold relic.    
    Taking into account the relations  among the annihilation    
cross--section at zero temperature $\langle\sigma_{\rm ann} v\rangle_0$ (relevant 
to the WIMP signal), 
the integral of $\langle\sigma_{\rm ann} v\rangle_{\rm FO}$    
from the present temperature up to the freeze-out temperature $T_f$ 
(relevant to the relic abundance) and the WIMP relic abundance    
$\Omega_{\chi} h^2$, by analytic arguments one derives that the quantity 
$\xi^2  \langle\sigma_{\rm ann} v\rangle_0$ has a maximum, {\it independent} of $m_{\chi}$.    
This maximum  is given by: $(\xi^2  \langle\sigma_{\rm ann} v\rangle_0)_{\rm max} = <\sigma_{\rm    
ann} v>_0|_{\eta = \eta'}$, where the values $\eta'$ are such that $(\Omega_{\chi} h^2)_{\eta =    
\eta'} =  (\Omega_{\rm CDM} h^2)_{\rm min}$ (see Ref. \cite{bdfs} for the details of the    
derivation). We disregard here special situations such as the one where 
co--annihilation would play a relevant role.    
    
An estimate for  $(\xi^2  \langle\sigma_{\rm ann} v\rangle_0)_{\rm max}$ can be derived analytically,    
if one employs the standard expansion in S and P waves:    
$\langle\sigma_{\rm ann} v\rangle \; \simeq \; \tilde{a} + \tilde{b}/x$,    
 for the thermally averaged product of    
the annihilation cross-section times the relative velocity of the self-interacting particles ($x=m_\chi/T$ where
$T$ is the temperature of the Universe), 
assuming that $\tilde{a} \geq |\tilde{b}|/(2 x_f) $ and that no substantial cancellations    
occur between the S and P terms ({\it i.e.}    
$\tilde{a} \gg |\tilde{b}|/(2 x_f) $), in case of negative values of $\tilde{b}$    
($x_f$ is the value    
of $x$ at freeze--out).    
    
By plugging in numbers (an important ingredient being the WMAP value    
$(\Omega_{\rm CDM} h^2)_{\rm min}$ = 0.092), one obtains \cite{bdfs}     
$(\xi^2  \langle\sigma_{\rm ann} v\rangle_0)_{\rm max} \simeq 3 - 5 \times 10^{-26} \,{\rm cm}^3 \cdot {\rm s}^{-1}$     
independently of $m_{\chi}$.     
This result, valid for a generic WIMP, is supported and strengthened     
by numerical analysis, in the case    
of relic neutralinos (see Fig. 2 of Ref. \cite{bdfs}), to:     
\begin{equation}    
(\xi^2  \langle\sigma_{\rm ann} v\rangle_0)_{\rm max} \simeq 3 \times 10^{-26} \, {\rm cm}^3 \cdot {\rm s}^{-1}\,.    
\label{eq:bound}    
\end{equation}    
Notice that, whereas the quantity $\xi^2  \langle\sigma_{\rm ann} v\rangle_0$ is limited from above    
by the bound of Eq. (\ref{eq:bound}),    
the quantity $\langle\sigma_{\rm ann} v\rangle_0$ is limited from below, since the cosmological bound    
 $\Omega_{\chi} h^2 \leq (\Omega_{\rm CDM} h^2)_{\rm max} \simeq$ 0.12 implies    
$\langle\sigma_{\rm ann} v\rangle_0 \gsim 9 \times 10^{-27}$ cm$^3 \cdot$ s$^{-1}$. However, for instance    
in the case    
of neutralinos,  $\langle\sigma_{\rm ann} v\rangle_0$ can take values much larger than its lower bound in    
sizable regions of the supersymmetric parameter space, though obviously preserving the upper limit    
of $\xi^2  \langle\sigma_{\rm ann} v\rangle_0$. Upper limits on $\langle\sigma_{\rm ann} v\rangle_0$  are only related     
to the particle physics properties of the WIMP candidate, and constrained by accelerator    
data and other laboratory precision measurement.  

Now we turn to the comparison of our bound of Eq. (\ref{eq:bound}) with the ones presented in 
Ref. \cite{hooper}. Notice that the analysis carried out in Ref. \cite{hooper} concerns 
WIMPs largely contributing to the galactic dark matter in a framework which includes non-thermal 
production or in general non-standard cosmology.  
 The upper bounds derived in  Ref. \cite{hooper} can be immediately converted into upper bounds in the scheme 
 we are discussing here (thermal WIMPs in a standard cosmological model) by simply substituting the quantity $\langle\sigma_{\rm ann} v\rangle_0$ on the vertical axis of Fig. 3 of Ref. \cite{hooper} by the quantity 
$\xi^2  \langle\sigma_{\rm ann} v\rangle_0$. 

Thus, we find that in the standard case of a thermal WIMP, the generic upper bound 
of Eq. (\ref{eq:bound}), based only on rescaling properties  and the low-energy
expansion, is more    
stringent than those derived in Ref. \cite{hooper}, which imply a number of specific 
astrophysical and particle--physics hypotheses.  The upper bounds 
of Ref. \cite{hooper} approch the upper limit of Eq. (\ref{eq:bound}) only in a narrow window with neutralinos  masses $\sim$     
100 GeV, when a Navarro--Frenk--White halo profile is assumed. { For a less steep density distribution,
like in the case of a cored profile, Ref. \cite{hooper} shows that the bound obtained from the synchrotron emission is significantly milder.} 

\begin{figure}[t] \centering
\includegraphics[width=1.1\columnwidth]{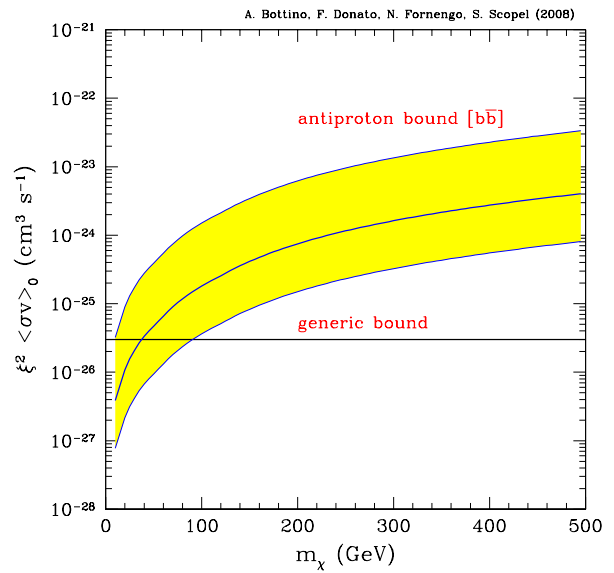}
\caption{\label{fig1} Generic upper bound of Eq.(\ref{eq:bound}) (horizontal solid line) and bounds from cosmic
antiproton searches (shaded area) for the quantity $\xi^2\langle\sigma_{\rm ann} v\rangle_0$. The generic upper bound applies
to thermal WIMPs. The antiproton bound is derived for the case of WIMP annihilation into a $b\bar b$ final state
\cite{noipbar}
and defined as the admissible excess in the available data \cite {pbardata} (for antiprotons kinetic energy of 0.23 GeV) over the standard background \cite{back} at 2$\sigma$. The upper,
median and lower lines refer to the maximal, median and minimal fluxes obtained by varying the
diffusion models parameters.}
\end{figure}

It is also worth noting that, as far as neutralinos are concerned, the bound    
$(\xi^2  \langle\sigma_{\rm ann} v\rangle_0)_{\rm max} \simeq 3 \times 10^{-26} \, {\rm cm}^3 \cdot {\rm s}^{-1}$    
is certainly more stringent than the ones derivable by present measurements of the cosmic positron    
spectrum    
and by the EGRET gamma ray spectrum. This simply because in these two cases the signals originated    
by neutralino annihilation require anyway a robust boost factor in order to fit the experimental data.    
Instead, for cosmic antiprotons, available experimental data can disallow a number    
of supersymmetric configurations for some values of the parameters of the astrophysical    
diffusion model \cite{lowind}.     
Note that this is at variance with what stated in Ref. \cite{hooper}, where the    
constraints from antiprotons are not considered and it is affirmed that    
the bound from synchrotron emission is more stringent than that    
provided by any other current indirect detection channel. 
As an example of the bound which can be obtained from the antiproton signal, we show in Fig. \ref{fig1}
the upper limit on $\xi^2\langle\sigma_{\rm ann} v\rangle_0$ obtained by considering the antiproton
flux from dark matter annihilation at kinetic energy $T_{\bar p} = 0.23$ GeV \cite{noipbar} and comparing it with the
current available data \cite{pbardata} and background determination \cite{back}. In Fig. \ref{fig1} we have considered the uncertainty in the calculation of the antiproton flux which originates from the galactic propagation.
For masses below 100 GeV and for a wide range of the astrophysical parameters involved in the diffusion
process, antiproton searches can set limits stronger than the generic bound of Eq. (\ref{eq:bound}).
In a more refined analysis, which takes into account the whole antiprotons energy spectrum,
the upper bound of Eq. (\ref{eq:bound})    
has been employed in Ref. \cite{bdfs}    
to establish that, in absence of a substantial clumpiness effect in the Galaxy    
\cite{lavalle},    
the cosmic antiproton spectrum is expected to be insensitive to neutralino annihilation for    
neutralino masses $\gsim$  200 GeV.   

Finally, we turn to some comments about the impact that our upper bound in Eq. (\ref{eq:bound})
can have on the interpretation of the WMAP haze in terms of thermal WIMPs. For this 
purpose a useful guide is provided by the lower frame of Fig. 3 in Ref. \cite{hfd},
   { where a density profile steeper than the NFW is adopted}.
   Firstly, by comparing the results there reported with the bounds derived in Ref. \cite{hooper}, 
   one realizes that large astrophysical and particle physics uncertainties affect the evaluation 
   of the haze effect. If we now wish to extract from the results of Ref. \cite{hfd}  the relevant 
   thermal WIMP candidates, we have to limit  the values of $\langle\sigma_{\rm ann} v\rangle_0$ from below,  
   since the cosmological upper bound  implies 
   $\langle\sigma_{\rm ann} v\rangle_0 \gsim 9 \times 10^{-27}$ cm$^3 \cdot$ s$^{-1}$. Taking into account 
   the bound of  Eq. (\ref{eq:bound}), one finds that, for most of the 
   annihilation modes, an interpretation of the WMAP haze 
  in terms of thermal WIMPs constituting the most part of dark matter favors relics 
   with a mass  $m_{\chi} \lsim$ 400 GeV. In the case of a direct annihilation into $e^+-e^-$,  
   larger values of the WIMP mass are accessible.
   
  To better constrain the model  in terms of thermal (dominant or subdominant) neutralinos,           
it would be interesting 
to carry out a detailed combined analysis of cosmic antiproton fluxes and of the WMAP haze effect,
 using, { for both type of signals}, the same diffusion model and density profile.

\acknowledgments We thank  G. Dobler, D.P. Finkerbeiner and D. Hooper for a correspondence clarifying 
some points of their papers. We acknowledge Research Grants funded jointly by    
the Italian Ministero dell'Istruzione, dell'Universit\`a e della    
Ricerca (MIUR), by the University of Torino and by the Istituto    
Nazionale di Fisica Nucleare (INFN) within the {\sl Astroparticle    
Physics Project}.

\end{document}